\documentclass[aps,pra,showpacs,floatfix,superscriptaddress]{revtex4}
\usepackage{graphicx}
\usepackage{subfigure}
\usepackage{color}
\usepackage{ulem}

\newcommand{\half}{\frac{1}{2}}
\begin{document}

\title{Entanglement Dynamics in Two-Qubit Open System Interacting with a
Squeezed Thermal Bath via Quantum Nondemolition interaction}

\author{Subhashish Banerjee}
\email{subhashish@cmi.ac.in}
\affiliation{Raman Research Institute, Bangalore- 560080, India}
\affiliation{Chennai Mathematical Institute, Padur PO, Siruseri 603103, India}
\author{V. Ravishankar}
\email{vravi@iitk.ac.in} 
\affiliation{Raman Research Institute, Bangalore- 560080, India}
\affiliation{Indian Institute of Technology, Kanpur, India} 
\author{R. Srikanth}
\email{srik@ppisr.res.in}
\affiliation{Poornaprajna Institute of Scientific Research, 
Bangalore- 560080, India}
\affiliation{Raman Research Institute, Bangalore- 560080, India}


\begin{abstract}
  We  analyze  the dynamics  of  entanglement  in  a two-qubit  system
  interacting  with an  initially squeezed  thermal environment  via a
  quantum nondemolition system-reservoir  interaction, with the system
  and  reservoir assumed  to be  initially separable.  We  compare and
  contrast the decoherence  of the two-qubit system in  the case where
  the qubits  are mutually  close-by (`collective regime')  or distant
  (`localized regime')  with respect to  the spatial variation  of the
  environment.   Sudden  death   of  entanglement   (as   quantified  by
  concurrence) is shown to occur  in the localized case rather than in
  the collective case, where  entanglement tends to `ring down'. 
  A consequence of the QND character of the interaction is that the
time-evolved fidelity of a Bell state never falls below $1/\sqrt{2}$,
a fact that is useful for quantum communication applications
like a quantum repeater. Using
  a  novel quantification of  mixed state  entanglement, we  show that
  there are noise regimes  where even though entanglement 
  vanishes,  the  state  is   still  available  for
  applications like  NMR quantum computation, because  of the presence
  of a pseudo-pure component.
\end{abstract} 


\maketitle

\section{Introduction}

Open quantum systems  are ubiquitous in the sense  that any system can
be thought  of as  being surrounded by  its environment  (reservoir or
bath) which influences its dynamics.  They provide a natural route for
discussing damping and dephasing. One of the first testing grounds for
open system  ideas was in quantum optics  \cite{wl73}. Its application
to other areas gained momentum  from the works of Caldeira and Leggett
\cite{cl83},   and  Zurek  \cite{wz93},   among  others.    The  total
Hamiltonian is  $H = H_S +  H_R + H_{SR}$  , where $S$ stands  for the
system,  $R$  for the  reservoir  and  $SR$  for the  system-reservoir
interaction.  The evolution  of the system of interest  $S$ is studied
taking into  account the  effect of its  environment $R$,  through the
$SR$    interaction    term,    making    the    resulting    dynamics
non-unitary. Depending upon  the system-reservoir ($S-R$) interaction,
open  systems can  be broadly  classified into  two  categories, viz.,
quantum non-demolition  (QND), which we consider  here, or dissipative
(cf. for  example Ref. \cite{ingold}).   A particular type  of quantum
nondemolition  (QND)  $S-R$  interaction   is  given  by  a  class  of
energy-preserving  measurements  in  which  dephasing  occurs  without
damping  the  system,  i.e.,  where  $[H_S, H_{SR}]  =  0$  while  the
dissipative systems  correspond to the case where  $[H_S, H_{SR}] \neq
0$ resulting in decoherence along with dissipation \cite{bg07}.

A class of observables that  may be measured repeatedly with arbitrary
precision,  with the  influence of  the measurement  apparatus  on the
system being confined strictly to the conjugate observables, is called
QND or back-action evasive observables \cite{bvt80, bk92, wm94, zu84}.
Such a measurement scheme was  originally introduced in the context of
the detection  of gravitational  waves \cite{ct80, bo96}.   The energy
preserving measurements, referred to above, form an important class of
such a general QND measurement scheme.

The  interest  in  the  relevance  of open  system  ideas  to  quantum
information has  increased in recent  times because of  the impressive
progress made, and the potential  for future progress (cf. for example
Ref. \cite{erika}),  on the experimental front in  the manipulation of
quantum states  of matter  towards quantum information  processing and
quantum communication.  Myatt {\it  et al.} \cite{myatt} and Turchette
{\it et  al.} \cite{turch} have  performed a series of  experiments in
which they induced  decoherence and decay by coupling  the atom (their
system-$S$) to  engineered reservoirs, in  which the coupling  to, and
the  state  of,  the  environment  are  controllable.   An  experiment
reported   in   Ref.     \cite{jb03}   demonstrated   and   completely
characterized a  QND scheme for making  a nondeterministic measurement
of  a single  photon  nondestructively using  only  linear optics  and
photo-detection of ancillary modes, to induce a strong nonlinearity at
the single  photon level.  The  dynamics of decoherence  in continuous
atom-optical QND measurements has been studied in \cite{vo98}.

Quantum entanglement is  the inherent property of a  system to exhibit
correlations, the physical basis being the non-local nature of quantum
mechanics  \cite{bell}, and hence  is a  property that  is exclusively
quantum  in  nature. Entanglement  plays  a  central  role in  quantum
information theory  \cite{nc}, in quantum  computation as in  the Shor
algorithm  \cite{shor},  and quantum  error  correction \cite{css}.  A
number  of  methods  have  been  proposed  for  creating  entanglement
involving trapped atoms \cite{beige, fy00, sm02}.

An important issue is to study how quantum entanglement is affected by
noise, which  can be thought of  as a manifestation of  an open system
effect  \cite{bp02}.   A  recent  experimental  investigation  of  the
dynamics  of   entanglement  with  a  continuous   monitoring  of  the
environment,   i.e.,  via  a   realization  of   quantum  trajectories
\cite{car93}, has been made in  \cite{am07}.  Here we study the effect
of noise on the entanglement generated between two spatially separated
qubits, by means of their interaction with the bath, which is taken to
be in an initial squeezed-thermal state \cite{bg07,sqgen}.  This is of
relevance to evaluate the  performance of two-qubit gates in practical
quantum information  processing systems.  The two  qubits are intially
uncorrelated.  With the advent  of time entanglement builds up between
them via their interaction with the bath but eventually gets destroyed
because of the quantum to  classical transition mediated by the noise.
In this  paper we study the effect  of noise generated by  a QND $S-R$
interaction.   The issue  of  a dissipative  noise  is taken  up in  a
separate work.

Since we are  dealing here with a two qubit  system which very rapidly
evolves into a mixed state,  a study of entanglement would necessarily
involve a measure of entanglement  for mixed states. Entanglement of a
bipartite system \cite{bz08}  in a pure state is  unambigious and well
defined.   However, mixed  state  entanglement (MSE)  is  not so  well
defined. Thus, although  a number of criteria such  as entanglement of
formation \cite{bd96,  ww98, mc05} and  negativity
\cite{fw89} exist, there  is a realization \cite {bd96}  that a single
quantity  is  inadequate  to  describe  MSE. This  was  the  principal
motivation  for   the  development  of  a  new   prescription  of  MSE
\cite{br08}  in which  it is  characterized not  
by a  number, but as  a probability density
function (PDF).   This  generalization provides
an exhaustive  and geometrical  characterization of entanglement  : by
exploring the  entanglement content in the  various subspaces spanning
the  two-qubit Hilbert  space.   The  known prescriptions
such as  concurrence and negativity emerge  as particular 
elements in the set  of parameters that characterize the
probability density function.  We will study entanglement
in the two-qubit  system using concurrence as well  as the probability
density function. 

The  plan of the  paper is  as follows.   In Sections  II and  III, we
develop our open  system model for the multi-qubit  dynamics under the
influence of a QND $S-R$ interaction.  Section II develops the general
dynamics  for a  multi-qubit system,  where the  qubits  are spatially
separated and  initially uncorrelated,  and the bath  is in  a general
squeezed-thermal state.  Section  III specializes these considerations
to  the  case  of two  qubits.   In  Section  IV,  we point  out  some
interesting symmetries  obeyed by  the two-qubit dynamics.   Section V
makes  a   brief  application  of  the  model   to  practical  quantum
communication, in particular, in the realization of a quantum repeater
\cite{chb96,bdcz98}.  In  Section VI, we  give a brief  description of
the  recently  developed  entanglement  measure  of  MSE  \cite{br08}.
Section VII deals with the entanglement analysis of the two-qubit open
system using the PDF as a  measure of entanglement. We also dwell upon
the usual  measure of  MSE, concurrence.  We  deal with  the scenarios
where  the  two  qubits   effectively  interact  via  localized  $S-R$
interactions, called the localized (independent) decoherence model, as
also  when  they  interact  collectively  with the  bath,  called  the
collective decoherence  model.  The usefulness  of the PDF  measure of
entanglement  is that  it allows  us to  demonstrate the  existence of
noise regimes  where even though  entanglement vanishes, the  state is
still available for applications like NMR quantum computation, because
of the  presence of a pseudo-pure  component.  In Section  VIII, as an
application of the PDF, we  make a brief discussion of the temperature
dependent effective  dynamics obeyed by  the two-qubit open  system in
the  collective  decoherence  regime.   In  Section IX,  we  make  our
conclusions.

\section{Multi -Qubit QND interaction with 
a Squeezed Thermal Bath}

We  consider the Hamiltonian,  describing the  QND interaction  of $L$
qubits with the bath as \cite{rj02, bg07, bgg07}
\begin{eqnarray}
H & = & H_S + H_R + H_{SR} \nonumber \\ & = & \sum\limits_{n=1}^L \hbar 
\varepsilon_n J^n_z+ \sum\limits_k \hbar \omega_k b^{\dagger}_k
b_k + \sum\limits_{n,k} \hbar J^n_z (g^n_k b^{\dagger}_k + 
g^{n*}_k b_k). \label{basich} 
\end{eqnarray} 
Here  $H_S$, $H_R$  and $H_{SR}$  stand  for the  Hamiltonians of  the
system,  reservoir  and  system-reservoir  interaction,  respectively.
$b^{\dagger}_k$, $b_k$ denote  the creation and annihilation operators
for the  reservoir oscillator of frequency  $\omega_k$, $g^n_k$ stands
for the coupling  constant (assumed to be position  dependent) for the
interaction  of the  oscillator  field with  the  qubit system and are
taken to be
\begin{equation}
g^n_k = g_k e^{-ik.r_n}, \label{coupling}
\end{equation}
where  $r_n$ is  the  qubit position.   Since  $[H_S, H_{SR}]=0$,  the
Hamiltonian (1) is of QND type. In the parlance of quantum information
theory,  the  noise  generated  is  called  the  phase  damping  noise
\cite{gp06, bg07}.

The  position dependence of  the coupling  of the  qubits to  the bath
(\ref{coupling}) helps to bring out the effect of entanglement between
qubits through the qubit separation: $ r_{mn} \equiv r_m - r_n$.  This
allows  for  a  discussion  of  the  dynamics  in  two  regimes:  (A).
localized decoherence where $k.r_{mn} \sim \frac{r_{mn}}{\lambda} \geq
1$   and   (B).    collective   decoherence   where   $k.r_{mn}   \sim
\frac{r_{mn}}{\lambda}  \rightarrow  0$. The  case  (B) of  collective
decoherence would arise  when the qubits are close  enough for them to
experience  the  same  environment,  or  when  the  bath  has  a  long
correlation  length  (set   by  the  effective  wavelength  $\lambda$)
compared to  the interqubit separation $r_{mn}$  \cite{rj02}.  Our aim
is to study the reduced dynamics  of the qubit system.  As in the case
of a  single qubit QND  interaction with bath \cite{bg07,  bgg07}, the
density matrix is  evaluated in the system eigenbasis  $|i_n \rangle =
|\pm \frac{1}{2}  \rangle$ (the  possible eigenstates of  $J^n_z$ with
eigenvalues  $j_n  =  \pm  \frac{1}{2}$).   The  system-plus-reservoir
composite is closed and hence  obeys a unitary evolution given, in the
interaction picture, by
\begin{equation}
\rho (t) = U_I (t) \rho (0) U^{\dagger}_I (t), 
\end{equation}
where
\begin{equation}
U_I (t) = {\cal T} e^{- (i/\hbar)\int_0^t dt^\prime H_I (t^\prime)},
\end{equation}
with $H_I (t) = e^{ i(H_S + H_R) t / \hbar} H_{SR} e^{- i(H_S + H_R) t
/ \hbar}$, and ${\cal T}$ denotes time ordering. Also
\begin{equation}
\rho (0) = \rho^s (0) \rho_R (0), \label{initial}
\end{equation}
i.e., we assume separable initial conditions.  Here
\begin{equation}
\rho^s (0) = \rho^s_1(0) \otimes \rho^s_2(0)\cdots \otimes
\rho^s_L(0), \label{sysinitial}
\end{equation}
is the initial state of the qubit system and the subscripts denote the
individual  qubits.  In  Eq.   (\ref{initial}), $\rho_R  (0)$  is  the
initial  density  matrix  of the  reservoir  which  we  take to  be  a
broadband squeezed thermal bath \cite{bg07,bgg07,gp06} given by
\begin{equation}
\rho_R(0) = S(\alpha, \Phi) \rho_{th} S^{\dagger} (\alpha, \Phi), \label{rhorin}
\end{equation}
where
\begin{equation}
\rho_{th} = \prod_k \left[ 1 - e^{- \beta \hbar \omega_k} 
\right] e^{-\beta \hbar \omega_k b^{\dagger}_k b_k} \label{rhoth}
\end{equation}
is the  density matrix  of the thermal  bath at temperature  $T$, with
$\beta \equiv 1/(k_B T)$, $k_B$ being the Boltzmann constant, and
\begin{equation}
S(\alpha_k, \Phi_k) = \exp \left[ \alpha_k \left( {b^2_k \over 2} e^{-2i 
\Phi_k} - {b^{\dagger 2}_k \over 2} e^{2i \Phi_k} \right) 
\right] \label{sqop}
\end{equation}
is  the  squeezing  operator   with  $\alpha_k$,  $\Phi_k$  being  the
squeezing parameters \cite{cs85}.

In order to obtain the reduced  dynamics of the system , we trace over
the reservoir  variables. The matrix  elements of the  reduced density
matrix in  the system  eigenbasis are obtained  for the  localized and
collective decoherence models as:

\subsection{Localized decoherence model}

\begin{equation}
\rho^s_{\{i_n,j_n\}} (t)  = \exp[i\{ \Theta_{\{i_n,j_n\}}^{\rm {lc}}  (t) - \Lambda_{\{i_n,j_n\}}^{\rm
{lc}} (t)\}] \exp[-\Gamma^{\rm  {lc}}_{\{i_n,j_n\} (\rm {sq})} (t)] \rho^s_{\{i_n,j_n\}}
(0). \label{ind}
\end{equation}
Here   $\rho^s_{\{i_n,j_n\}}    (t)$   stands   for    $\langle   i_L,
i_{L-1},...,i_1|  {\rm Tr}_R{\rho^s (t)}|j_L,  j_{L-1},...,j_1 \rangle$
and    the   symbol    ${\{i_n,j_n\}}$    stands   collectively    for
${i_1,j_1;i_2,j_2;...;i_L,j_L}$.   The  superscript  {\it  lc}  is  to
indicate that  these expressions  are for the  localized decoherence
model and the  subscript {\it sq} indicates that the  bath starts in a
squeezed thermal  initial state.  As  seen from the  expressions given
below,          $\Theta_{\{i_n,j_n\}}^{\rm          {lc}}$         and
$\Lambda_{\{i_n,j_n\}}^{\rm {lc}}$ are independent of the bath initial
conditions  and   are  given  in  the  continuum   limit  (assuming  a
quasi-continuous bath spectrum) by
\begin{equation}
\Theta_{\{i_n,j_n\}}^{\rm  {lc}}  (t)  = 2\int\limits^{\infty}_0  d\omega  I(\omega)
S(\omega,t)  \sum^L_{\stackrel{m=1,n=2}{(m \neq  n)}} (i_m  i_n  - j_m
j_n) \cos(\omega t_s),
\label{thetain}
\end{equation}
\begin{equation}
\Lambda_{\{i_n,j_n\}}^{\rm  {lc}} (t)  =  2\int\limits^{\infty}_0 d\omega  I(\omega)
C(\omega,t)   \sum^L_{(m    \neq   n)}   i_m   j_n
\sin(\omega t_s).
\label{lambdain}
\end{equation}
In  the above  equations,  $I(\omega)$ is  the  bath spectral  density
given in  terms of the  system (qubits) and  bath coupling
constant  as $I(\omega)=  \sum\limits_{k} \delta  (\omega  - \omega_k)
g^2_k$, which for the Ohmic case considered here has the
form
\begin{equation}
I(\omega) = {\gamma_0 \over \pi} \omega e^{-\omega/\omega_c}, 
\label{ohmic} 
\end{equation}
where $\gamma_0$ and $\omega_c$ are two bath parameters. Also
\begin{equation}
S(\omega, t) = \frac{\omega t - \sin(\omega t)}{\omega^2}, \label{sin}
\end{equation}
and
\begin{equation}
C(\omega, t) = \frac{1 - \cos(\omega t)}{\omega^2}. \label{cos}
\end{equation}
In Eqs.  (\ref{thetain}) and  (\ref{lambdain}), $\omega t_s  \equiv k.
r_{mn}$  \cite{rj02}, where  $t_s$ is  the transit  time
introduced  in  order  to  express  the system-bath  coupling  in  the
frequency    domain.  $\Gamma^{lc}_{sq}    (t)$   in
Eq. (\ref{ind}) is given as
\begin{eqnarray}
\Gamma^{\rm  {lc}}_{\{i_n,j_n\} (\rm {sq})}  (t) &=&  \int\limits^{\infty}_0 d\omega
I(\omega)  \coth(\frac{\beta  \hbar  \omega}{2}) \nonumber\\  &\times&
\left[\cosh(2\alpha)C(\omega,  t)  \Big\{\sum^L_{m=1}(i_m   -  j_m)^2  +  2
\sum^L_{\stackrel{m=1,n=2}{(m   \neq  n)}}   (i_m-  j_m)   (i_n-  j_n)
\cos(\omega        t_s)\Big\}\right.\nonumber\\       &-&       \left.
\frac{2}{\omega^2}\sin^2(\frac{\omega         t}{2})         \sinh(2\alpha)
\Big\{\cos(\omega(t-2a))[\sum^L_{m=1}(i_m    -    j_m)^2   \cos(\omega
t_{corr1})\right.         \nonumber\\       &+&        \left.        2
\sum^L_{\stackrel{m=1,n=2}{(m   \neq  n)}}   (i_m-  j_m)   (i_n-  j_n)
\cos(\omega t_{corr2})]+  \sin(\omega(t-2a))[\sum^L_{m=1}(i_m - j_m)^2
\sin(\omega   t_{corr1})   \right.     \nonumber\\   &+&   \left.    2
\sum^L_{\stackrel{m=1,n=2}{(m   \neq  n)}}   (i_m-  j_m)   (i_n-  j_n)
\sin(\omega t_{corr2})]\Big\}\right], \label{gammain}
\end{eqnarray}
where we  have defined two  new time scales $\omega t_{corr1}  \equiv 2k.r_m$
and  $\omega t_{corr2}   \equiv  k.(r_n  +   r_m)$  which  are  due   to  the
non-stationary effects  introduced by the squeezed  thermal bath. Here
we have for simplicity taken the squeezed bath parameters as
\begin{eqnarray} 
\cosh \left( 2\alpha(\omega) \right) & = & \cosh (2\alpha),~~ \sinh 
\left( 2\alpha (\omega) \right) = \sinh (2\alpha), \nonumber\\ \Phi 
(\omega) & = & a\omega, \label{sqpara} 
\end{eqnarray} 
where $a$ is a constant depending upon the squeezed bath.

\subsection{Collective decoherence model}

The reduced density matrix is given by 
\begin{equation}
\rho^s_{\{i_n,j_n\}} (t) = \exp[i\{  \Theta_{\{i_n,j_n\}}^{\rm {col}} (t) - \Lambda_{\{i_n,j_n\}}^{\rm
{col}}    (t)\}]     \exp[-\Gamma^{\rm    {col}}_{\{i_n,j_n\} (\rm{sq})}    (t)]
\rho^s_{\{i_n,j_n\}} (0). \label{col}
\end{equation}
 The superscript $\rm {col}$ is to indicate that these expressions are
for  the collective  decoherence model  and the  subscript  $\rm {sq}$
indicates that  the bath starts  in a squeezed thermal  initial state.
As in  the case of  localized decoherence, $\Theta^{\rm  {col}}$ and
$\Lambda^{\rm {col}}$  are independent of the  bath initial conditions
and are given in the continuum limit (assuming a quasi-continuous bath
spectrum) by
\begin{equation}
\Theta^{\rm  {col}}  (t)  = \int\limits^{\infty}_0  d\omega  I(\omega)
S(\omega,t) \left[(\sum^L_{m=1} i_m)^2 - (\sum^L_{m=1} j_m)^2\right],
\label{thetacol}
\end{equation}
\begin{equation}
\Lambda^{\rm {col}} (t) = 0.
\label{lambdacol}
\end{equation}
The bath spectral  density $I(\omega)$ is as in  Eq. (\ref{ohmic}).  In
Eq. (\ref{col}), $\Gamma^{\rm {col}}_{\{i_n,j_n\} (\rm{sq})} (t)$ is
\begin{eqnarray}
\Gamma^{\rm {col}}_{\{i_n,j_n \}(\rm{sq})} (t) &=&  \int\limits^{\infty}_0 d\omega
I(\omega) \coth(\frac{\beta \hbar \omega}{2}) \left[\cosh(2\alpha)C(\omega,
t)  [\sum^L_{m=1}(i_m  -   j_m)]^2  \right.   \nonumber\\  &-&  \left.
\frac{2}{\omega^2}\sin^2(\frac{\omega         t}{2})         \sinh(2\alpha)
\Big\{\cos(\omega(t-2a))[\sum^L_{m=1}(i_m    -    j_m)^2   \cos(\omega
t_{corr1})\right.         \nonumber\\       &+&        \left.        2
\sum^L_{\stackrel{m=1,n=2}{(m   \neq  n)}}   (i_m-  j_m)   (i_n-  j_n)
\cos(\omega t_{corr2})] + \sin(\omega(t-2a))[\sum^L_{m=1}(i_m - j_m)^2
\sin(\omega   t_{corr1})   \right.     \nonumber\\   &+&   \left.    2
\sum^L_{\stackrel{m=1,n=2}{(m   \neq  n)}}   (i_m-  j_m)   (i_n-  j_n)
\sin(\omega t_{corr2})]\Big\}\right], \label{gammacol}
\end{eqnarray}
All  the  other  terms  are   as  defined  above.   On  comparing  Eq.
(\ref{gammacol})  with   (\ref{gammain}),  we  find   that  the  terms
proportional to $\sinh(2\alpha)$, arising from the non-stationarity of
the  squeezed   bath,  are  same  while  the   terms  proportional  to
$\cosh(2\alpha)$   differ  from  each   other.   For   the  collective
decoherence model,  $\omega t_s \equiv  k.  r_{mn} \equiv 0$,  but the
two  time-scales  coming from  the  non-stationary  components of  the
squeezed  thermal bath,  i.e.,  $\omega t_{corr1}  \equiv 2k.r_m$  and
$\omega t_{corr2} \equiv k.(r_n  + r_m)$ are both non-zero, indicative
of correlations induced between the qubits by the bath squeezing.  For
the case of  zero bath squeezing, both the  Eqs.  (\ref{gammacol}) and
(\ref{gammain}) reduce to their corresponding values for the case of a
thermal bath \cite{rj02}.

\section{Two qubit interaction}

Here we specialize the  general considerations of the previous section
to the case of two qubits. 

\subsection{Localized decoherence model \label{sec:3a}}

The reduced density  matrix is a specialization of  Eq. (\ref{ind}) to
the case of two qubits, say a and b. Here $\rho^s_{\{i_n,j_n\}}(t)$ would
be  $\rho^s_{\{i_a,j_b\}} (t)$  which represents  $\langle i_a,  i_b| {\rm
Tr}_R{\rho^s (t)}|j_a, j_b \rangle$, where the states $|i_a\rangle$ or
$|i_b\rangle$   have   eigenvalues   $\pm   \frac{1}{2}$.    We   will
collectively represent the two-particle index $ab$ by a single 4-level
index according to the following scheme:
$$  -\frac{1}{2},-\frac{1}{2}   \equiv  0,~~  -\frac{1}{2},\frac{1}{2}
\equiv       1,~~      \frac{1}{2},-\frac{1}{2}       \equiv      2,~~
\frac{1}{2},\frac{1}{2}  \equiv  3.  $$  
Thus  there  will be  sixteen
elements of  the density matrix,  which we enumerate below.   They are
seen to satisfy the symmetries 
\begin{equation}
\rho^s_{32} (t) = \rho^{\ast s}_{23} (t) = \rho^s_{01} (t)
= \rho^{\ast s}_{10} (t),
\label{ina}
\end{equation}
where $\ast$ in the  superscript indicates complex conjugation, and of
course the first and last  equality follow from the hermiticity of the
density  operator.  In  the Eqs.   (\ref{ina}), $\Theta_{\{i_n,j_n\}}^{\rm
{lc}} (t)$,  $\Lambda_{\{i_n,j_n\}}^{\rm {lc}}  (t)$ can be  obtained from
the   Eqs.   (\ref{thetain}),   (\ref{lambdain}),   respectively,  and
$\Gamma^{\rm   {lc}}_{\{i_n,j_n\}   (\rm   {sq})}   (t)$  from     Eq.
(\ref{gammain}) and are given by
\begin{equation}
\Theta^{\rm  {lc}}_{32}(t) = \Theta^{\rm {lc}}_{01}(t) 
=  \int\limits^{\infty}_0 d\omega I(\omega)  S(\omega,t) \cos(\omega
t_s),
\label{theta1}
\end{equation}
\begin{equation}
\Lambda^{\rm  {lc}}_{32}(t)  =
\Lambda^{\rm  {lc}}_{01}(t) 
=  -\int\limits^{\infty}_0 d\omega I(\omega)  C(\omega,t) \sin(\omega
t_s),
\label{lambda1}
\end{equation}
\begin{equation}
\Theta^{\rm  {lc}}_{23}(t)  =
\Theta^{\rm  {lc}}_{10}(t)  =
-\Theta^{\rm  {lc}}_{32}(t) =
-\Theta^{\rm  {lc}}_{01}(t),
\label{theta2}
\end{equation}
\begin{equation}
\Lambda^{\rm  {lc}}_{23}  (t)  =
\Lambda^{\rm {lc}}_{10}(t) =
-\Lambda^{\rm {lc}}_{32}(t) =
-\Lambda^{\rm {lc}}_{01}(t),
\label{lambda2}
\end{equation}
and
\begin{eqnarray}
\Gamma^{\rm  {lc}}_{\rm {sq}}  (t) &=&  \int\limits^{\infty}_0 d\omega
I(\omega) \coth(\frac{\beta \hbar \omega}{2}) \left[\cosh(2\alpha)C(\omega,
t)-     \frac{2}{\omega^2}\sin^2(\frac{\omega     t}{2})     \sinh(2\alpha)
\Big\{\cos(\omega(t-2a))  \cos(\omega t_{corr1}^{(1)})  \right.  \nonumber\\
&+& \left.  \sin(\omega(t-2a)) \sin(\omega t_{corr1}^{(1)})\Big\}\right],
\label{gamma1}
\end{eqnarray}
for all the  above combinations. In the above  equations, $\omega t_s$
stands  for  $k\cdot  r_{ab}$  while  $\omega  t_{corr1}^{(1)}  \equiv
2k\cdot  r_b$.  Interestingly,  for the  above cases,  the correlation
time $\omega  t_{corr2} \equiv  k.(r_a + r_b)$  is absent.  It  can be
seen that
\begin{eqnarray}
\rho^s_{aa} (t) &=&
\rho^s_{aa} (0), ~~~~(a = 0,1,2,3) \label{inb},
\end{eqnarray}
from which follows  that the population remains unchanged.   This is a
consequence of QND nature of the $S-R$ interaction. Also,
\begin{eqnarray}
\rho^s_{21} (t) = \rho^{\ast s}_{12} (t) = \rho^s_{12} (t),\nonumber\\
\rho^s_{30} (t) = \rho^{\ast s}_{03} (t) = \rho^s_{03} (t),  \label{inc}
\end{eqnarray}
i.e.,  these components are  purely real.   In   Eqs.  (\ref{inc}),
$\Theta^{\rm  {lc}} (t)$,  $\Lambda^{\rm {lc}}  (t)$  and $\Gamma^{\rm
  {lc}}_{\rm {sq}} (t)$ are given by
\begin{equation}
\Theta^{\rm  {lc}} (t) = 0 = \Lambda^{\rm  {lc}} (t), \label{theta3}
\end{equation}
and 
\begin{eqnarray}
\Gamma^{\rm {lc}}_{ {\rm sq},30}    (t)    &=&
\Gamma^{\rm {lc}}_{ {\rm sq},03}    (t)    = \nonumber\\    
&=&   \int\limits^{\infty}_0    d\omega   I(\omega)
\coth(\frac{\beta \hbar  \omega}{2}) \left[2\cosh(2\alpha)C(\omega, t)  [1 +
\cos(\omega   t_s)]  -   \frac{2}{\omega^2}\sin^2(\frac{\omega  t}{2})
\sinh(2\alpha)       \right.        \nonumber\\       &\times&       \left.
\Big\{\cos(\omega(t-2a))   [\cos(2k\cdot r_a)   +   \cos(2k\cdot r_b) + 2\cos(k\cdot [r_a+r_b])] \right.  \nonumber \\
&+& \left. \sin(\omega(t-2a))
[\sin(2k\cdot r_a)   +   \sin(2k\cdot r_b) + 2\sin(k\cdot [r_a+r_b])] 
\Big\}\right],
\label{gamma2}
\end{eqnarray}
\begin{eqnarray}
\Gamma^{\rm                                                  {lc}}_{\rm
{sq}, 21}    (t)    &=&
\Gamma^{\rm                                                  {lc}}_{\rm
{sq}, 12}            (t)
\nonumber\\    &=&      \int\limits^{\infty}_0    d\omega   I(\omega)
\coth(\frac{\beta \hbar  \omega}{2}) \left[2\cosh(2\alpha)C(\omega, t)  [1 -
\cos(\omega   t_s)]  -   \frac{2}{\omega^2}\sin^2(\frac{\omega  t}{2})
\sinh(2\alpha)       \right.        \nonumber\\       &\times&       \left.
\Big\{\cos(\omega(t-2a))   [\cos(2k\cdot r_a)   +   \cos(2k\cdot r_b) - 2\cos(k\cdot [r_a+r_b])]   \right.   
\nonumber\\  &+&   \left.   \sin(\omega(t-2a))
[\sin(2k\cdot r_a)   +   \sin(2k\cdot r_b) - 2\sin(k\cdot [r_a+r_b])]\Big\}\right].
\label{gamma3}
\end{eqnarray}
Thus we  see that the  Eqs. (\ref{gamma2}), (\ref{gamma3}),  depend on
both $2 k\cdot r_a$ and $2 k\cdot r_b$,
and $\omega t_{corr2}$ which is as defined above. Further,
\begin{equation}
\rho^s_{31} (t) = \rho^{\ast s}_{13} (t) = \rho^s_{02} (t) = 
\rho^{\ast s}_{20} (t),
\label{rhoind}
\end{equation}
where $\ast$  in the superscript indicates 
complex conjugation.  In
 Eqs.  (\ref{rhoind}), $\Theta^{\rm {lc}} (t)$, $\Lambda^{\rm {lc}}
(t)$ are
\begin{equation}
\Theta^{\rm  {lc}}_{31}(t) = \Theta^{\rm {lc}}_{02}(t) 
=  \int\limits^{\infty}_0 d\omega I(\omega)  S(\omega,t) \cos(\omega
t_s),
\label{theta4}
\end{equation}
\begin{equation}
\Lambda^{\rm  {lc}}_{31}(t)  =
\Lambda^{\rm  {lc}}_{02}(t) 
=  \int\limits^{\infty}_0 d\omega I(\omega)  C(\omega,t) \sin(\omega
t_s),
\label{lambda4}
\end{equation}
\begin{equation}
\Theta^{\rm  {lc}}_{13}(t)  =
\Theta^{\rm  {lc}}_{20}(t)  =
-\Theta^{\rm  {lc}}_{31}(t) =
-\Theta^{\rm  {lc}}_{02}(t),
\label{theta5}
\end{equation}
\begin{equation}
\Lambda^{\rm  {lc}}_{13}  (t)  =
\Lambda^{\rm {lc}}_{20}(t) =
-\Lambda^{\rm {lc}}_{31}(t) =
-\Lambda^{\rm {lc}}_{02}(t),
\label{lambda5}
\end{equation}
and
\begin{eqnarray}
\Gamma^{\rm  {lc}}_{\rm {sq}}  (t) &=&  \int\limits^{\infty}_0 d\omega
I(\omega) \coth(\frac{\beta \hbar \omega}{2}) \left[\cosh(2\alpha)C(\omega,
t)-     \frac{2}{\omega^2}\sin^2(\frac{\omega     t}{2})     \sinh(2\alpha)
\Big\{\cos(\omega(t-2a))  \cos(\omega t_{corr1}^{(2)})  \right.  \nonumber\\
&+& \left.  \sin(\omega(t-2a)) \sin(\omega t_{corr1}^{(2)})\Big\}\right],
\label{gamma5}
\end{eqnarray}
for all the  above combinations. In the above  equations, $\omega t_s$
stands  for $k\cdot  r_{ab}$  while $\omega  t_{corr1}^{(2)} \equiv  2k\cdot
r_a$.   Interestingly,  for  the  above cases,  the  correlation  time
$\omega  t_{corr2}  \equiv  k.(r_a   +  r_b)$  is  absent.   The  Eqs.
(\ref{ina}), (\ref{inb}), (\ref{inc}) and (\ref{rhoind}) cover all the
density matrices for the two-qubit localized decoherence model.
It can be shown from these results  that with the increase in temperature,
as also  evolution time  $t$ and bath  squeezing $\alpha$,  the system
becomes more mixed and hence looses its purity.

\subsection{Collective decoherence model}

The reduced density  matrix is a specialization of  Eq. (\ref{col}) to
the case of two qubits, say a and b. The notations are as before.
\begin{equation}
\rho^s_{32} (t) = \rho^{\ast s}_{23} (t) = \rho^s_{01} (t) = 
\rho^{\ast s}_{10} (t),
\label{cola}
\end{equation}
where $\ast$  in the superscript indicates  complex conjugation.  In
the Eqs. (\ref{cola}),  $\Theta^{\rm {col}} (t)$, $\Lambda^{\rm {col}}
(t)$   ($=  0$)   are  obtained   from  the   Eqs.   (\ref{thetacol}),
(\ref{lambdacol}),  respectively  and  $\Gamma^{\rm {col}}_{\rm  {sq}}
(t)$ from Eq.  (\ref{gammacol}). They are given by
\begin{equation}
\Theta^{\rm  {in}}_{32}(t) = \Theta^{\rm in}_{01}(t) 
=  \int\limits^{\infty}_0 d\omega I(\omega)  S(\omega,t),
\label{thetacol1}
\end{equation}
\begin{equation}
\Theta^{\rm  {col}}_{23}(t)  =
\Theta^{\rm  {col}}_{10}(t)  =
-\Theta^{\rm  {col}}_{32}(t) =
-\Theta^{\rm  {col}}_{01}(t),
\label{thetacol2}
\end{equation}
and $\Gamma^{\rm {col}}_{\rm  {sq}} (t)$ is as in Eq. (\ref{gamma1})
for all the cases in Eq. (\ref{cola}), with $\omega  t_{corr1}$ and
$\omega  t_{corr2}$ as defined there. As before,
\begin{eqnarray}
\rho^s_{aa} (t) &=&
\rho^s_{aa} (0), ~~~~(a = 0,1,2,3) \label{colb}.
\end{eqnarray}
This  is indicative  of  QND  nature of  the  $S-R$ interaction  which
preserves the population. Also,
\begin{eqnarray}
\rho^s_{21} (t) = \rho^{\ast s}_{12} (t) = \rho^s_{12} (t),\nonumber\\
\rho^s_{30} (t) = \rho^{\ast s}_{03}  (t) = \rho^s_{03} (t).  \label{colc}
\end{eqnarray}

In  the Eqs.   (\ref{colc}), $\Theta^{\rm  {col}}  (t)$, $\Lambda^{\rm
{col}} (t)$ and $\Gamma^{\rm {col}}_{\rm {sq}} (t)$ are given by
\begin{equation}
\Theta^{\rm  {col}} (t) = 0 = \Lambda^{\rm  {col}} (t), \label{thetacol3}
\end{equation}
and 
\begin{eqnarray}
\Gamma^{\rm                                                  {col}}_{\rm
{sq}, 30}    (t)    &=&
\Gamma^{\rm                                                  {col}}_{\rm
{sq}, 03}            (t)
\nonumber\\    &=&     \int\limits^{\infty}_0    d\omega   I(\omega)
\coth(\frac{\beta \hbar  \omega}{2}) \left[4 \cosh(2\alpha)C(\omega, t)  
-   \frac{2}{\omega^2}\sin^2(\frac{\omega  t}{2})
\sinh(2\alpha)       \right.        \nonumber\\       &\times&       \left.
\Big\{\cos(\omega(t-2a))   [\cos(2k\cdot r_a)   +   \cos(2k\cdot r_b) 
+ 2\cos(k\cdot [r_a+r_b])]   \right.   \nonumber\\  
&+&   \left.   \sin(\omega(t-2a))
[\sin(2k\cdot r_a)   +   \sin(2k\cdot r_b) + 2\sin(k\cdot [r_a+r_b])]\Big\}\right],
\label{gammacol2}
\end{eqnarray}
\begin{eqnarray}
\Gamma^{\rm                                                 {col}}_{\rm
{sq}, 21}    (t)    &=&
\Gamma^{\rm                                                 {col}}_{\rm
{sq}, 12}            (t)
\nonumber\\ &=&  - 2 \int\limits^{\infty}_0  \frac{d\omega} {\omega^2}
I(\omega)  \coth(\frac{\beta   \hbar  \omega}{2})  \sin^2(\frac{\omega
t}{2})   \sinh(2\alpha)   \nonumber\\   &\times&   \left[\cos(\omega(t-2a))
[\cos(2k\cdot r_a)   +   \cos(2k\cdot r_b) 
- 2\cos(k\cdot [r_a+r_b])] \right.  \nonumber\\
&+& \left.   \sin(\omega(t-2a)) [\sin(2k\cdot r_a)   +   \sin(2k\cdot r_b) - 2
\sin(k\cdot [r_a+r_b])]\right].
\label{gammacol3}
\end{eqnarray}
It is  interesting to note from  Eqs.  (\ref{colc}), (\ref{thetacol3})
and (\ref{gammacol3}), that for the case of a purely thermal bath with
zero  bath  squeezing,  $\Gamma^{\rm   {col}}_{\rm  {sq},  21}  (t)  =
\Gamma^{\rm {col}}_{\rm {sq}, 12} (t)  = 0$, thereby implying that for
these cases,  the corresponding density  matrix elements do  not decay
even though they are interacting with  the bath.  Also, since in a QND
$S-R$ interaction, the diagonal terms $\rho_{1,1}$ and $\rho_{2,2}$ do
not   change,  this   implies  that   any  state   $\alpha|1\rangle  +
\beta|2\rangle$ in the subspace span $\{|1\rangle,|2\rangle\}$ remains
invariant, thereby leading to a {\it decoherence-free subspace}.  
Further,
\begin{equation}
\rho^s_{31} (t) = \rho^{\ast s}_{13} (t) = \rho^s_{02} (t) = 
\rho^{\ast s}_{20} (t),
\label{cold}
\end{equation}
where $\ast$  in the superscript indicates  complex conjugation.
In Eq.  (\ref{cold}), $\Theta^{\rm {col}}(t)$, $\Lambda^{\rm {col}} (t)$
$(=0)$ are
\begin{equation}
\Theta^{\rm  {col}}_{31}(t) = \Theta^{\rm col}_{02}(t) 
=  \int\limits^{\infty}_0 d\omega I(\omega)  S(\omega,t),
\label{thetacol4}
\end{equation}
\begin{equation}
\Theta^{\rm  {col}}_{13}(t)  =
\Theta^{\rm  {col}}_{20}(t)  =
-\Theta^{\rm  {col}}_{31}(t) =
-\Theta^{\rm  {col}}_{02}(t),
\label{thetacol5}
\end{equation}
and $\Gamma^{\rm {col}}_{\rm {sq}}  (t)$ is as in Eq.  (\ref{gamma1}),
with $2k\cdot  r_b \rightarrow 2k\cdot r_a$.   The Eqs.  (\ref{cola}),
(\ref{colb}),  (\ref{colc})  and (\ref{cold})  cover  all the  density
matrices for the two-qubit collective decoherence model.
As  with the localized case, it  can  be seen  that with  the
increase in temperature, as also evolution time $t$ and bath squeezing
$\alpha$, the system becomes more mixed and hence looses its purity.

\section{Symmetries in the dynamical system}

In this section  we consider the two qubit evolution,  developed in the
previous  section   from  the  point  of  view   of  some  fundamental
symmetries.   This  enables us  to  view  the  dynamics from  a  fresh
perspective and is also interesting from its own point of view.

Employing  the  two-particle  index  notation  used  in  the  previous
section, we  find that the  transformation connecting the  initial and
final density operations can be given by the following operation
\begin{equation}
\rho_{ab}(t) = {\cal L}_{ab}(t) \rho_{ab}(0).
\end{equation}
The non-trivial  aspect of the dynamics that  this relation represents
is  that   here  ${\cal   L}$  represents,  not   a  matrix,   but  a
two-dimensional array, and the multiplication is done element-wise.

The most general array ${\cal L}$ that satisfies this property, following only
from the hermiticity of $\rho(0)$ and $\rho(t)$ is:
\begin{equation}
{\cal L} = \left( \begin{array}{cccc}
1 	& c_1 	& c_2 	& c_3 \\
c_1^*	& 1	& c_4	& c_5 \\
c_2^*	& c_4^*	& 1	& c_6 \\
c_3^*	& c_5^* & c_6^*	& 1
\end{array} \right)
\label{eq:L}
\end{equation}

However,  further constraints on  the structure  of ${\cal  L}$ appear
because  the  dynamical  evolution  due to  QND  interaction  respects
spin-flip  symmetry (see  Eq.  (\ref{spinflip})  below), which  is for
example (given for clarity, in the single-qubit notation):
\begin{equation}
{\cal L}_{\half,-\half;\half,\half} = {\cal L}_{-\half,\half;-\half,-\half}.
\end{equation} 
This has the effect that $c_3$ and $c_4$ are real, 
which we denote by $r_1$ and $r_2$ respectively.
Further
$c_2 = c_5^*$ and $c_1 = c_6^*$. These are seen by noting that:
\begin{eqnarray}
c_3 &\equiv& {\cal L}_{e,g} = {\cal L}^*_{g,e} = {\cal L}_{g,e} = {\cal L}^*_{e,g} \equiv c_3^*, \nonumber \\
c_4 &\equiv& {\cal L}_{s,a} = {\cal L}^*_{a,s} = {\cal L}_{a,s} = {\cal L}^*_{s,a} \equiv c_4^*, \nonumber \\
c_2 &\equiv& {\cal L}_{e,a} = {\cal L}^*_{a,e} = {\cal L}_{g,s} = {\cal L}^*_{s,g} \equiv c_5^*, \nonumber \\
c_1 &\equiv& {\cal L}_{e,s} = {\cal L}^*_{s,e} = {\cal L}_{g,a} = {\cal L}^*_{a,g} \equiv c_6^*,
\end{eqnarray}
where the first and third equalities in each equation follow from hermiticity.
Accordingly, Eq. (\ref{eq:L}) can be rewritten as:
\begin{equation}
{\cal L} = \left( \begin{array}{cccc}
1 	& c_1 	& c_2 	& r_1 \\
c_1^*	& 1	& r_2	& c_2^* \\
c_2^*	& r_2	& 1	& c_1^* \\
r_1	& c_2 & c_1	& 1
\end{array} \right),
\label{eq:LL}
\end{equation}

Consider   the  operator  $\hat{\cal   L}$  corresponding   to  ${\cal
L}_{j,k}$, defined by:
\begin{equation}
\hat{\cal L} = \sum_{j,k} {\cal L}_{j,k} |j\rangle\langle k|,~~~~ (j,k = 0,1,2,3).
\end{equation}
The spin-flip symmetry can be represented by
\begin{equation}
\Sigma \hat{\cal L} \Sigma^\dag = \hat{\cal L},~~~~\Sigma=\sigma_x \otimes \sigma_x . \label{spinflip}
\end{equation}
Since $\Sigma = (-i\sigma_x) \otimes (i\sigma_x)$, the above spin-flip
symmetry may be described as a rotational symmetry, with angle $\pi/2$
(resp. $3\pi/2$)  in the first  (resp. second) qubit  coordinate about
the $x$-axis.

\section{An application to quantum communication: quantum 
repeaters}

We now make an application  of the two-qubit reduced dynamics obtained
from   QND  system-reservoir   interaction  to   a   quantum  repeater
\cite{bdcz98},  used for  quantum communication  over  long distances.
The efficiency of quantum communication over long distances is reduced
due to the effect of noise,  which can be considered as a natural open
system effect.  For distances much longer than the coherence length of
a noisy  quantum channel, the  fidelity of transmission is  usually so
low  that  standard purification  methods  are  not  applicable. In  a
quantum repeater set-up, the  channel is divided into shorter segments
that  are purified  separately and  then  connected by  the method  of
entanglement swapping, which is the quantum teleportation \cite{qtele}
of entanglement. This  method can be much more  efficient than schemes
based on quantum error correction, as it makes explicit use of two-way
classical   communication.   The   quantum   repeater  system   allows
entanglement purification  over arbitrary long  channels and tolerates
errors  on the  percent level.  It requires  a polynomial  overhead in
time,   and  an   overhead  in   local  resources   that   grows  only
logarithmically with the length of the channel.

Here we  consider the effect  of noise, introduced by  imperfect local
operations that constitute the  protocols of entanglement swapping and
purification \cite{chb96}, on such a  compound channel, and how it can
be kept below  a certain threshold.  The noise  process studied is the
one  obtained   from  the  two-qubit   reduced  dynamics  via   a  QND
interaction,   instead  of  the   depolarizing  noise   considered  in
\cite{bdcz98}.  A detailed study of  the effect of the two-qubit noise
on  the performance  of a  quantum repeater  is underway  and  will be
reported  elsewhere.  Here  we  treat  this problem  in  a  simplified
fashion, and  study the  applicability and efficiency  of entanglement
purification protocols in the situation of imperfect local operations.

A quantum  repeater involves the  two tasks of  entanglement swapping,
involving  Bell-state  measurements,  and  entanglement  purification,
involving CNOT  gates. The Bell-state measurement  may be equivalently
replaced   by   a  CNOT   followed   by   a  projective   single-qubit
measurement. In  entanglement swapping, two  distant parties initially
not  sharing entanglement  with each  other, but  sharing entanglement
separately  with  a third  party,  become  entangled  by virtue  of  a
multi-partite  measurement by  the  third party  on  the latter's  two
halves  of  entanglement.    Entanglement  purification  involves  two
parties employing local  operations and classical communication (LOCC)
to  improve the  fidelity $F$  of Einstein-Podolsky-Rosen  (EPR) pairs
they share,  with respect to  a maximally entangled state.   The local
operations  involve  two-qubit  gates  such  as  the  CNOT  operation,
followed by single qubit measurement,  and a possible discarding of an
EPR pair. Provided $F>0.5$, and  at the cost of losing shared (impure)
entanglement,  the  two  parties  can  increase the  fidelity  of  the
remaining shared entanglement to
\begin{equation}
F^\prime = \frac{F^2 + [(1-F)/3]^2}{F^2 + [2F(1-F)/3] + (5/9)(1-F)^2},
\label{eq:bennett}
\end{equation}
where  $F$ and  $F^\prime$  are, respectively,  the  input and  output
fidelities (with respect to a Bell state) 
of  the  entanglement  purification protocol  proposed  by
Bennett {\it et al.} \cite{chb96}.

\begin{figure}
\includegraphics[width=7.0cm]{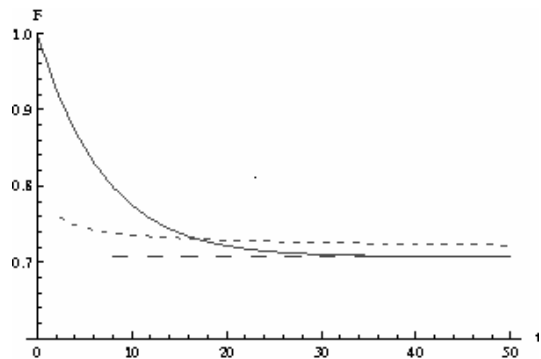}
\caption{Fidelity  (with  respect  to  the Bell  state  $|00\rangle  -
  |11\rangle$)  as   a  function  of  time  for   a  two-qubit  system
  interacting  with its environment  via a  QND interaction.  The bold
  curve corresponds  to a finite  temperature and no  squeezing ($T=4,
  \alpha=0$); the  small-dashed curve corresponds  to zero temperature
  but  finite  squeezing  ($T=0,  \alpha=2$); the  large-dashed  curve
  corresponds to  finite temperature and  squeezing ($T=4, \alpha=2$).
  For  sufficiently large  time, the  fidelity  approaches $1/\sqrt{2}
  \approx 0.707$.}
\label{fig:repeater-00}
\end{figure}

We consider two repeaters in a realistic situation where they are well
separated and hence lie in the localized regime of our model.  If they
initially     share    a     Bell    state     $|\psi\rangle    \equiv
(1/\sqrt{2})(|00\rangle -  |11\rangle) \equiv (1/\sqrt{2})(|g\rangle -
|e\rangle)$, a QND interaction  will asymptotically drive the state to
the   maximally  mixed   state  with   support   in  span$\{|g\rangle,
|e\rangle\}$,  i.e.,   $\rho_r  \equiv  (1/2)(|g\rangle\langle   g|  +
|e\rangle\langle   e|)$.   The   asymptotic  fidelity   is   given  by
$F=\sqrt{\langle\psi|\rho_r|\psi\rangle}  = \frac{1}{\sqrt{2}} \approx
0.707$,  a  pattern  evident  from  Figure  \ref{fig:repeater-00}.   A
similar result of course can be given for the other Bell states. Since
this value exceeds 0.5, pairs of  qubits that start out in a maximally
entangled  state can  always  be distilled  via  the quantum  repeater
scheme.  In all  the figures in this article,  we consider the initial
state to  be an  equal superposition state,  which can be  obtained by
applying   $H   \otimes   H$    on   the   state   $|0\rangle   \equiv
|-\frac{1}{2},-\frac{1}{2}\rangle$,   where   $H$   is  the   Hadamard
transformation.  The figure shows  that environmental  squeezing, like
temperature, impairs  fidelity, and  can thus not  be used  to counter
thermal effects. This concurrent behavior of squeezing and temperature
for  QND type  of interactions  is  mirrored also  in phase  diffusion
\cite{bgg07} and the evolution of geometric phase \cite{gp06}.

\section{Characterization of Mixed State Entanglement through 
a Probability  Density Function}

As mentioned  in the  introduction, it  is  important to
determine the entanglement in the two  qubit system, if it is to be of
utility  in  quantum information  processing.  There  is, however,  no
straightforward way of determining the entanglement when the system is
in  a  mixed  state since,  as  is  well  known, entanglement,  as  an
observable,   cannot   be    represented   by   a   linear   hermitian
operator. Indeed, it  is impossible to capture the  information on the
entanglement  in  a mixed  state  by  a  single parameter  \cite{bd96},
notwithstanding the  fact that  useful benchmarks such  as concurrence
and  negativity   exist  .   Thus  ,  for   instance,  negativity  and
concurrence are not relative monotones.

For this reason, we employ  a recently proposed description of MSE via
a  probability density  function.  We also  determine the  concurrence
(negativity for a state can never exceed its concurrence), and show it
along with the PDF, for comparison, and to display the extent to which
it captures the information on the entanglement in the state.

Here  we  briefly recapitulate  the  characterization  of mixed  state
entanglement  (MSE) through a  PDF as  developed in  \cite{br08}.  The
basic idea  is to express  the PDF of  entanglement of a  given system
density matrix (in this case, a  two-qubit) in terms of a weighted sum
over the PDF's of projection operators spanning the full Hilbert space
of the system density matrix.  Consider first a system in
a  state which is  a projection  operator of  dimension $d$.  The pure
states correspond to $d=1$, and  the completely mixed states to $d=4$.
The PDF of a system in a  state which is a projection operator $\rho =
\frac{1}{M}\Pi_M$ of rank $M$ is defined as:
\begin{equation}
{\cal P}_{\Pi_M}({\cal E}) = \frac{\int d{\cal H}_{\Pi_M}\delta
({\cal E}_{\psi} - {\cal E})}{\int d{\cal H}_{\Pi_M}}, \label{entproject}
\end{equation}
where  $\int  d{\cal H}_{\Pi_M}$  is  the  volume  measure for  ${\cal
  H}_{\Pi_M}$, which  is the subspace spanned by  $\Pi_M$.  The volume
measure is  determined by the  invariant Haar measure  associated with
the  group of automorphisms  of $\int  d{\cal H}_{\Pi_M}$,  modulo the
stabilizer group of the reference state generating ${\cal H}_{\Pi_M}$.
Thus for  a one dimensional  projection operator, representing  a pure
state,  the  group of  automorphisms  consists  of  only the  identity
element and  the PDF is  simply given by  the Dirac delta.  Indeed, if
$\rho= \Pi_1  \equiv \vert \psi  \rangle \langle \psi \vert$,  the PDF
has the  form ${\cal  P}_{\rho}({\cal E}) =  \delta ({\cal E}  - {\cal
E}_{\psi})$
thereby resulting  in the description  of pure state  entanglement, as
expected, by a single number.  The entanglement density of a system in
a general  mixed state  $\rho$ is  given by resolving  it in  terms of
nested projection operators with appropriate weights as
\begin{eqnarray} 
\rho &=& (\lambda_1 - \lambda_2)\Pi_1 + (\lambda_2 - \lambda_3)\Pi_2
+ .......(\lambda_{N-1} - \lambda_N)\Pi_{N-1} + \lambda_N \Pi_N  \nonumber\\
&\equiv& \sum_{M=1}^N \Lambda_M \Pi_M, \label{nested}
\end{eqnarray}
where  the projections  are  $\Pi_M =\sum_{j=1}^M|\psi_j\rangle\langle
\psi_j|$,  with $M  = 1,2,...,N$  and the  eigenvalues  $\lambda_1 \ge
\lambda_2  \ge  ....$,  i.e.,   the  eigenvalues  are  arranged  in  a
non-increasing fashion. Thus the PDF for the entanglement of $\rho$ is
given by
\begin{equation}
{\cal P}_{\rho} ({\cal E}) =  \sum_{M=1}^N \omega_M {\cal P}_{\Pi_M}
({\cal E}), \label{PDF}
\end{equation}
where  the weights  of  the respective  projections ${\cal  P}_{\Pi_M}
({\cal E})$ are given by  $\omega_M = \Lambda_M/\lambda_1$.  For a two
qubit system, the density matrix  would be represented as a nested sum
over  four projection  operators, $\Pi_1$,  $\Pi_2$,  $\Pi_3$, $\Pi_4$
corresponding  to one,  two, three  and four  dimensional projections,
respectively, with  $\Pi_1$ corresponding to a pure  state and $\Pi_4$
corresponding  to a  a uniformly  mixed state,  is a  multiple  of the
identity  operator.   The most  interesting  structure  is present  in
$\Pi_2$,  the two-dimensional  projection, which  is  characterized by
three parameters,  viz. ${\cal  E}_{cusp}$, the entanglement  at which
the PDF  diverges, ${\cal E}_{max}$, the  maximum entanglement allowed
and  ${\cal P}_2 ({\cal  E}_{max})$, the  PDF corresponding  to ${\cal
  E}_{max}$. The three dimensional projection $\Pi_3$ is characterized
by the parameter ${\cal E}_{\perp}$, which parametrizes a discontunity
in  the  entanglement  density  function  curve.   By  virtue  of  the
convexity of  the sum over  the nested projections  (\ref{nested}), it
can be seen  that the concurrence of any state $\rho$  is given by the
inequality   ${\cal  C}_{\rho}   \leq  (\lambda_1   -  \lambda_2){\cal
  C}_{\Pi_1} +  (\lambda_2 - \lambda_3){\cal  C}_{\Pi_2}$.  Thus while
the  concurrence  for  a  three  and four  dimensional  projection  is
identically  zero, through the  PDF one  is able  to make  a statement
about the entanglement content of states which span 
these spaces. 
 Also, as pointed out
in \cite{br08},  in the case  of NMR quantum  computation, concurrence
and negativity are zero, whereas the PDF is able to elucidate the role
of entanglement utilized by the NMR operations. These features as well
as the fact that the  PDF (\ref{PDF}) enables us to study entanglement
of  a  physical state  by  exploiting  the  richness inherent  in  the
subspaces  spanned  by the  system  Hilbert  space  makes the  PDF  an
attractive statistical and geometric characterization of entanglement.
We provide an explicit illustration of this in the next 
section.

\section{Entanglement analysis}

In this section, we will  study the development of entanglement in the
two qubit  system, both  for the localized  as well as  the collective
decoherence model. Recall that
concurrence \cite{ww98} is defined as
\begin{equation} 
{\cal C} = \max(0, \sqrt{\lambda_1} - \sqrt{\lambda_2} - \sqrt{\lambda_3}-
\sqrt{\lambda_4}), \label{concur}
\end{equation}
where ${\lambda_i}$ are the eigenvalues of the matrix 
\begin{equation}
R = \rho \tilde{\rho}, \label{concur1}
\end{equation}
with $\tilde{\rho} = \sigma_y \otimes \sigma_y \rho^* \sigma_y \otimes
\sigma_y$ and $\sigma_y$ is the usual Pauli matrix. ${\cal C}$ is zero
for unentangled states and one for maximally entangled states.  In the
above expression, it is implicitly assumed that $\rho$ is expressed in
a seperable basis.

\begin{figure}
\subfigure[]{\includegraphics[width=7.0cm]{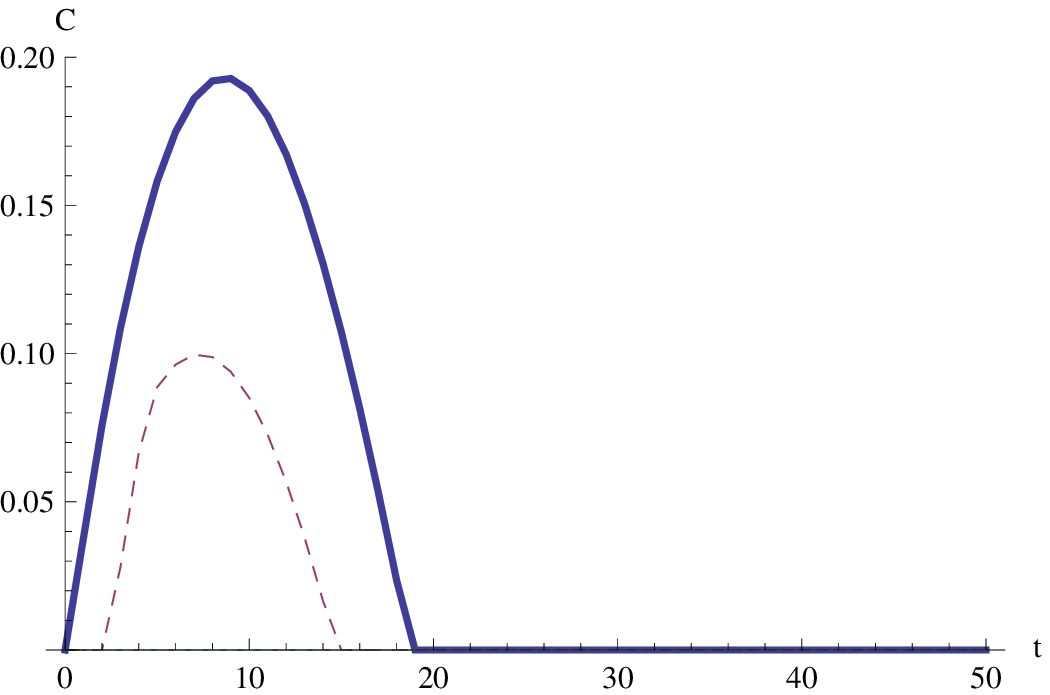}} 
\hfill
\subfigure[]{\includegraphics[width=7.0cm]{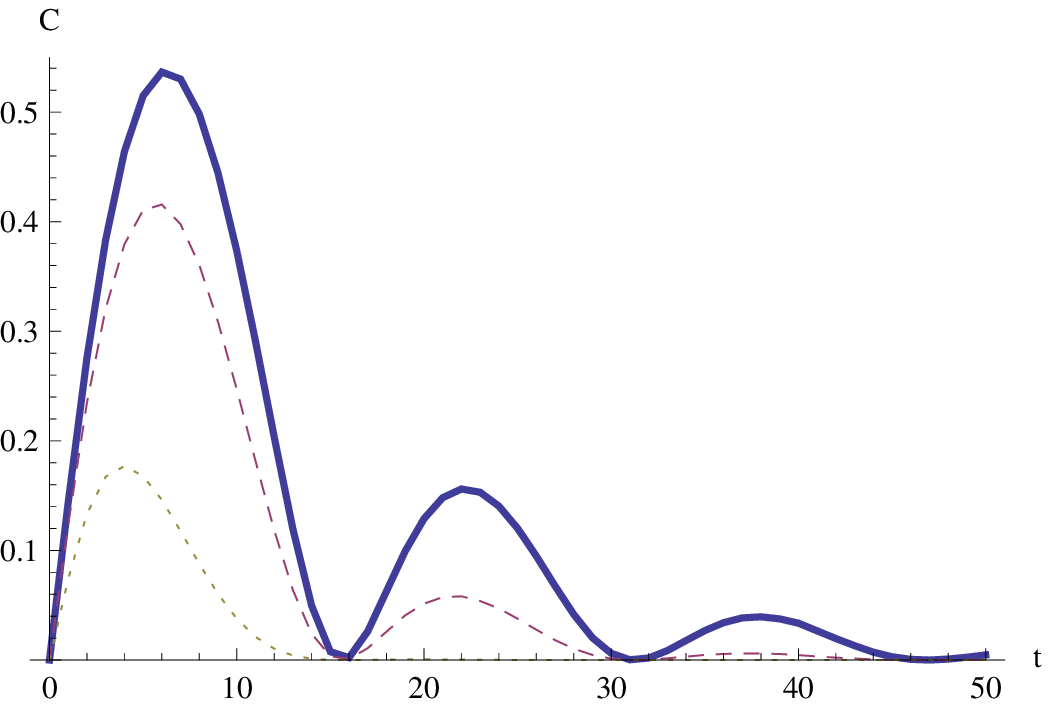}}
\caption{Concurrence ${\cal  C}$ (\ref{concur}) as a  function of time
  of evolution $t$ at $T  = 5.0$ and bath squeezing parameter $\alpha$
  (\ref{sqpara})  equal to  $0$,  $0.5$, $1.0$,  corresponding to  the
  bold, dashed and dotted  curves, respectively.  Figure (a) refers to
  the localized  decoherence model and (b)  the collective decoherence
  model.}
\label{fig:concurrence}
\end{figure}

In  figure  (\ref{fig:concurrence}   (a)),  we  plot  the  concurrence
(\ref{concur})  with respect  to time  for the  case of  the localized
decoherence  model, while  figure (\ref{fig:concurrence}  (b)) depicts
the temporal  behavior of  concurrence for the  collective decoherence
model,  for different bath  squeezing parameters.   In all  figures in
this  article,  $k r_{12}$  is  set equal  to  1.1  for the  localized
decoherence model  and 0.05 for  the collective model.  It  is clearly
seen  from  the  figures  that  the  two  qubit  system  is  initially
unentangled, but with time there is a build up of entanglement between
them  as  a result  of  their interaction  with  the  bath.  Also  the
entanglement  builds up  more  quickly in  the collective  decoherence
model when compared  to the localized model.  This  is expected as the
effective  interaction  between the  two  qubits  is  stronger in  the
collective  case.  Another  interesting feature  that can  be inferred
from   figure  (\ref{fig:concurrence}   (a))  is   the   phenomena  of
entanglement birth and death  \cite{ye09} in the localized decoherence
model. Figure (\ref{fig:concurrence}  (b)) exhibits entanglement death
followed by revival, in the  collective decoherence model. It is clear
from the figures that  bath squeezing retards the dynamical generation
of  entanglement.  However,  interestingly,  it is  observed that  the
disentanglement  time  is  the   same  for  different  bath  squeezing
parameters, as in figure  (\ref{fig:concurrence} (b)), while it varies
for the  independent decoherence model,  figure (\ref{fig:concurrence}
(a)).   This  indicates a  kind  of  robustness  of the  phenomena  of
disentanglement  with respect  to  bath squeezing,  in the  collective
regime.


There  have  been a  number  of  investigations  in the  phenomena  of
entanglement  sudden death  and  revival. In  \cite{ye06}  a study  of
entanglement sudden  death and revival  was made between  two isolated
atoms  each  in its  own  lossless  Jaynes-Cummings  cavity, while  in
\cite{ye07} the evolution of  entanglement was studied via information
exchange  between  subsystems  rather  than decoherence.   Thus  these
studies revealed  features of the dynamics  of entanglement generation
in  the absence  of  decoherence.  In  another  study \cite{ft06}  was
revealed the  interesting effect that  irreversible spontaneous decay,
due to  interaction with  a vacuum  bath, can have  on the  revival of
entanglement between two qubits with the collective decoherence regime
being most conducive to the revival of entanglement.  Even though this
work involved  a dissipative system-bath  interaction, this conclusion
is  supported  here,  for  QND  interactions,  as  the  generation  of
entanglement  is seen  to be  much  more effective  in the  collective
regime  when   compared  to  the  independent  one.    The  effect  of
non-Markovian  influences, due  to a  dissipative interaction,  on the
dynamics of entanglement between two qubits was studied in \cite{bc08}
for  the  localized  decoherence  model  and in  \cite{mp08}  for  the
collective regime. Here our study concentrates on QND interactions and
localized as  well as  collective regimes are  treated under  a common
footing.

\begin{figure}
\subfigure[]{\includegraphics[width=7.0cm]{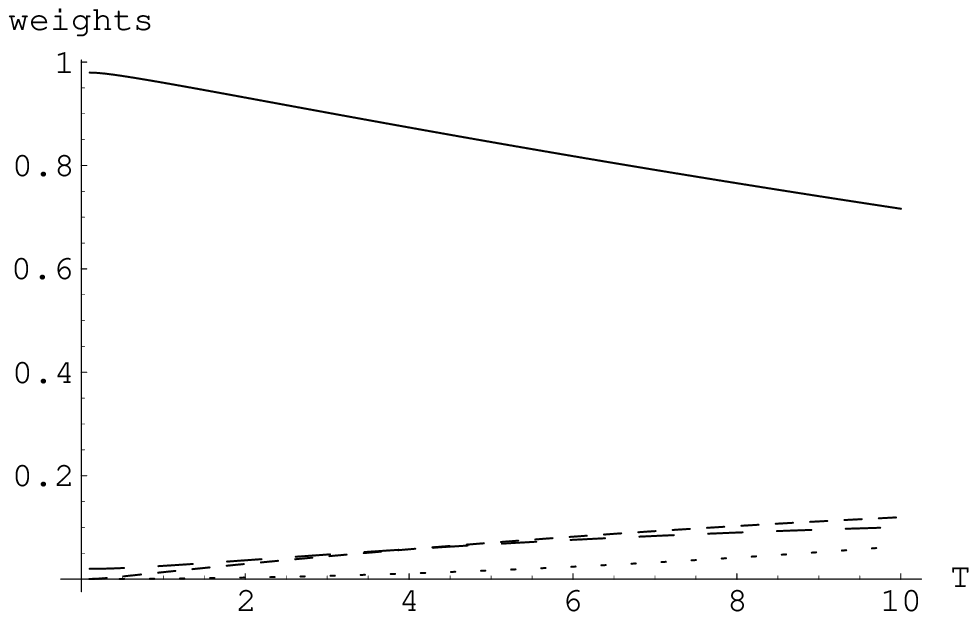}} 
\hfill
\subfigure[]{\includegraphics[width=7.0cm]{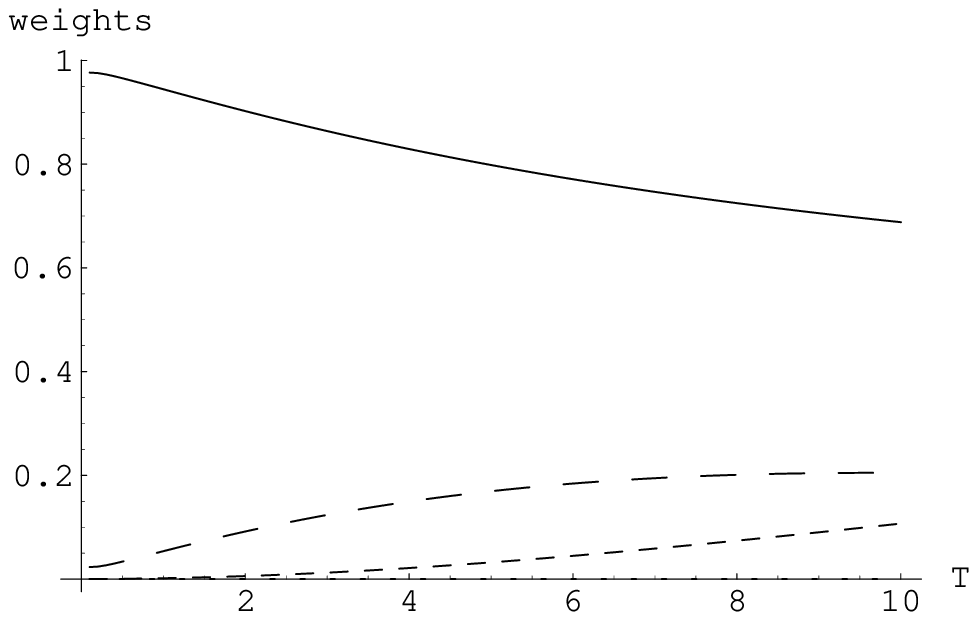}}
\caption{The  weights  (\ref{PDF})  as  a  function of  $T$,  with  an
  evolution  time  $t  =  5$  and bath  squeezing  parameter  $\alpha$
  (\ref{sqpara}) equal to $0.2$.  Figure (a) refers to the localized
  decoherence model and (b) the collective decoherence model.  In both
  the figures,  the bold curve  corresponds to the  weight $\omega_1$,
  while the large-dashed, small-dashed and dotted curves correspond to
  the weights $\omega_2$, $\omega_3$ and $\omega_4$, respectively.}
\label{fig:weights}
\end{figure}

Now we take  up the issue of entanglement from  the perspective of the
PDF as  in Eq.  (\ref{PDF}).   In figures (\ref{fig:weights}  (a)) and
(b),  we  plot  the  weights $\omega_1$,  $\omega_2$,  $\omega_3$  and
$\omega_4$ (\ref{PDF}) of the entanglement densities of the projection
operators of  the various subspaces  which span the two  qubit Hilbert
space with respect to $T$ for the localized and collective decoherence
models, respectively.   As can  be seen from  both the figures  , with
increase in temperature $T$, the weight $\omega_1$, depicting the pure
state component monotonically decreases, while the other weights start
from zero  at $T=0$ and  increase.  Eventually, the  weight $\omega_4$
depicting  a maximally  mixed  state would  be  expected to  dominate,
though for the parameter range used  in the plots, this feature is not
seen.  This feature  of the dynamics of the  reduced two-qubit system,
specially  in the  case of  the collective  decoherence model,  has an
interesting application  which will be discussed in  detail in Section
VIII  where it will  be seen  to obey  an {\it  effective} temperature
dependent  Hamiltonian, bringing out  the persistence  of entanglement
even  at   finite  temperatures.   In  the  case   of  the  collective
decoherence  model,  the  weights  $\omega_2$ and  $\omega_3$  have  a
greater  growth  than  that   for  the  localized  decoherence  model,
depicting the greater entanglement development in the collective model
as   is  also   borne  out   by  the   concurrence  plots   in  figure
(\ref{fig:concurrence}).

As  explained in Section  VI, the  characterization of  MSE for  a two
qubit system via  the PDF involves the distribution  functions of four
projection operators, $\Pi_1$, $\Pi_2$, $\Pi_3$, $\Pi_4$ corresponding
to  one, two,  three and  four dimensional  projections, respectively.
These will be represented here as ${\cal P}_1 ({\cal E})$, ${\cal P}_2
({\cal  E})$, ${\cal  P}_3 ({\cal  E})$ and  ${\cal P}_4  ({\cal E})$,
respectively. Also, as discussed  above, ${\cal P}_4 ({\cal E})$ would
be universal for the two  qubit density matrices and would involve the
Haar measure on $SU(4)$ \cite{br08,tbs02}.

Consideration of  the ${\cal P}_2  ({\cal E})$ and ${\cal  P}_3 ({\cal
  E})$  density functions for  some representative  states of  the two
qubit system, both for the localized as well as collective decoherence
models,  enables us  to  compare the  entanglement  in the  respective
subspaces  of the  system Hilbert  space.  The  details of
these density  functions for  different parameters, pertaining  to the
two-qubit reduced dynamics, have been presented in \cite{entqnd09}.

\begin{figure}
\subfigure[]{\includegraphics[width=7.0cm]{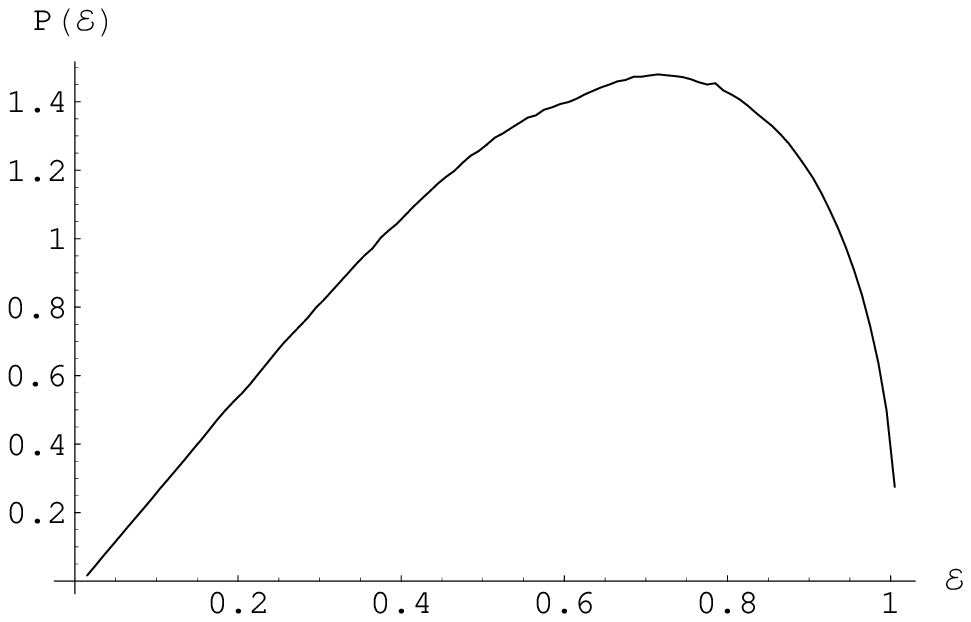}}
\hfill
\subfigure[]{\includegraphics[width=7.0cm]{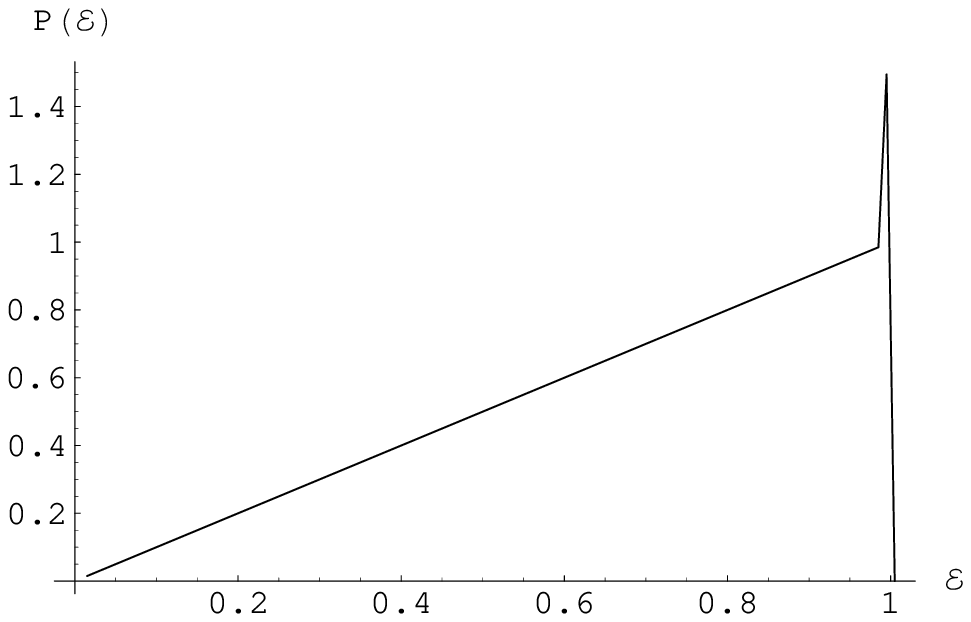}}
\caption{The  full density function  ${\cal P}({\cal  E})$ (\ref{PDF})
  with respect to the entanglement ${\cal E}$ for an evolution time $t
  = 10.0$, $T  = 50.0$ and bath squeezing  parameter $\alpha$ equal to
  $0.2$. Figure  (a) refers to  the localized decoherence  model and
  (b) to the collective decoherence model.}
\label{fig:PfullT50}
\end{figure}

Figures  (\ref{fig:PfullT50}  (a))  and  (b)  give  the  full  density
function  ${\cal   P}(\cal  E)$  for  the   localized  and  collective
decoherence  models, respectively,  with a  bath evolution  time  $t =
10.0$ and $T = 50.0$.   For these conditions, the value of concurrence
(\ref{concur})  is 0,  which would  indicate a  complete  breakdown of
entanglement. This would be expected  as with the increase in the bath
temperature  $T$,  the  effect  of  entanglement  would  be  destroyed
quickly.  This is  partially borne out by the fact  that for this case
${\cal   C}_{\Pi_2}   =   0$.     However,   as   seen   from   figure
(\ref{fig:PfullT50} (b)), the PDF  for the full density function still
exhibits a  rich entanglement  structure, coming principally  from the
contributions  from the  one  and three  dimensional projections.   In
contrast,   figure  (\ref{fig:PfullT50}   (a)),   for  the   localized
decoherence  model, exhibits  the  Haar measure  on  $SU(4)$ and  thus
represents a maximally mixed state.

\begin{figure}
\subfigure[]{\includegraphics[width=7.0cm]{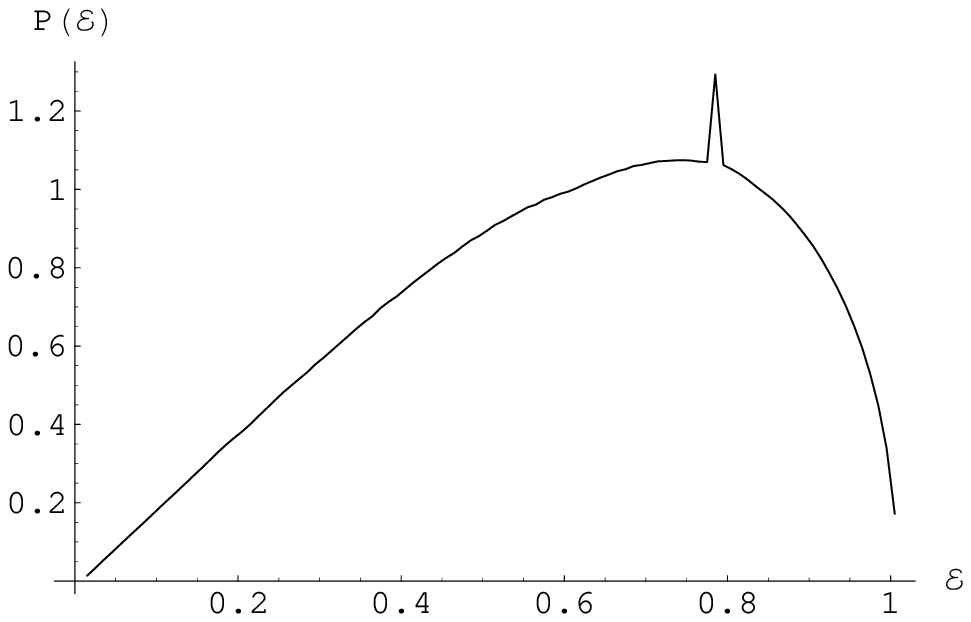}}
\hfill
\subfigure[]{\includegraphics[width=7.0cm]{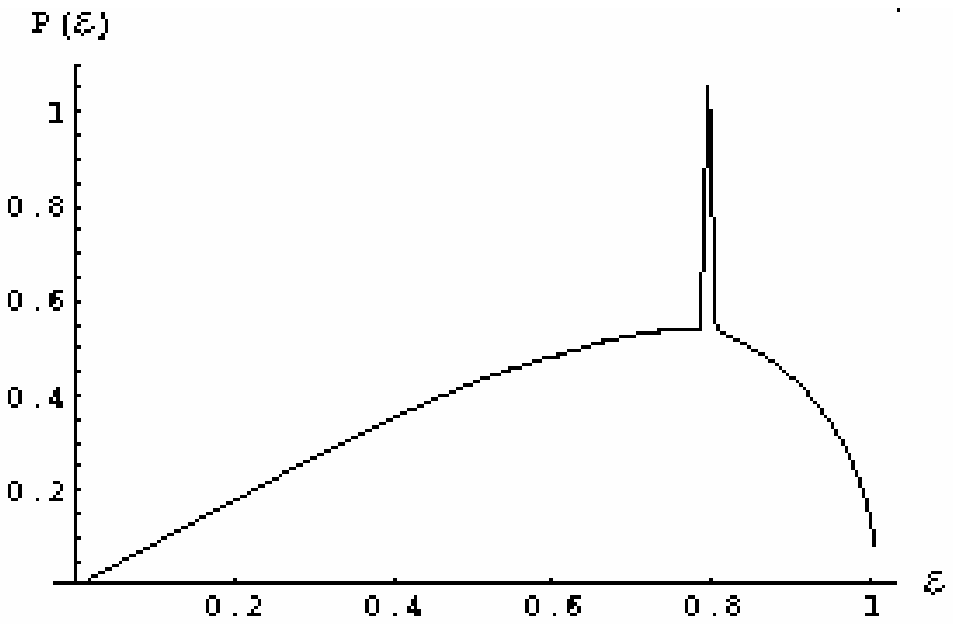}}
\caption{The  full density function  ${\cal P}({\cal  E})$ (\ref{PDF})
  with  respect to  the entanglement  ${\cal E}$  for  the localized
  decoherence model with an evolution time  $t = 10.0$, $T = 20.0$ and
  bath squeezing parameter $\alpha$ equal to $0.2$ (figure (a)) and $0$
(figure (b)).}
\label{fig:PfullT20}
\end{figure}

Figures  (\ref{fig:PfullT20})  represents  the full  density  function
${\cal  P}(\cal  E)$  for  the  localized decoherence  model  with  an
evolution time  $t =  10.0$, $T =  20.0$ and bath  squeezing parameter
$\alpha$ equal to  $0.2$, figure (a), and equal to  $0$ in figure (b).
This case  is interesting since it  is analogous to  that discussed in
\cite{br08}  for NMR  quantum computation  where concurrence  would be
zero,  and  the  excess  of  entangled  states  over  the  unpolarized
background  (exhibited by  the  uniform distribution  coming from  the
density   function  $\Pi_4$,   related  to   the   fourth  dimensional
projection) is  exploited as a resource allowing  for non-trivial gate
operations, thus depicting pseudopure states over the four dimensional
background, with  the excess being the ``deviation  density matrix". A
comparison  of   the  two  figures   shows  that  the   generation  of
entanglement is  greater for  the case of  zero bath squeezing,  as in
figure (b), when compared to the case of finite bath squeezing, figure
(a).

\section{ Effective temperature dependent dynamics in the collective 
decoherence model: a brief discussion}

In a QND  $S-R$ interaction, the reduced density  matrix of the system
does  not approach a  unique distribution  asymptotically \cite{bg07}.
It turns out that the  PDF for  the full
density  function (for  the collective  decoherence model)  exhibits a
rich entanglement structure, coming principally from the contributions
from  the one  and  three dimensional  projections  which carry  equal
weights. This feature  is seen to persist for  higher temperatures and
evolution  times,  for  the  collective decoherence  model,  with  the
weights of the  subspaces spanned by the four  projection operators of
the PDF  remaining intact.   From this emerges  the fact that  for the
collective decoherence model, studied here,  as the effect of the bath
on  the system  increases, the  PDF  instead of  becoming uniform,  as
expected, gets  distributed between the  subspaces spanned by  the one
and three  dimensional projection  operators suggesting a  tendency of
the system  to resist  randomization.  We  may ask  if the
description  of   the  mixed  state  entanglement  in   terms  of  the
probability  density function  which we  employ here  can throw,  as a
nontrivial  application, light on  the effective  dynamics of  the two
qubit system.  It has been  seen in earlier
studies  that   such  a  state  of  affairs would  be
encountered if  the effect  of the  bath is not  a counterpart  of the
collision term  (in a Boltzmann equation),  but is more  like a Vlasov
term,  causing  long   range  mean  field  contributions  \cite{ce94}.
We analyse our system in detail below. 

Indeed, from the numerical  results, it is
not difficult to see that the effect of the
bath  can be  mapped to  a $T$  dependent effective  hamiltonian whose
energy eigenvalues scale with  temperature.  The eigenstates are given
by the standard Bell states  with the ground state being $|B_1 \rangle
= \frac{1}{\sqrt{2}} (|01 \rangle + |10 \rangle)$ while the orthogonal
singlet state  ($|B_4 \rangle = \frac{1}{\sqrt{2}} (|01  \rangle - |10
\rangle)$) is  the highest energy state, and  is practically decoupled
(with no population).  The next  excited state is degenerate, with two
Bell  states ($|B_2 \rangle  = \frac{1}{\sqrt{2}}  (|00 \rangle  - |11
\rangle)$,  $|B_3  \rangle =  \frac{1}{\sqrt{2}}  (|00  \rangle +  |11
\rangle)$) spanning  the two dimensional  subspace.  Thus
it  follows  that the  effective temperature  dependent
hamiltonian is given by
$$
H_{\rm eff} = \sum^{4}_{i=1} E_i(\beta) |B_i\rangle\langle B_i|, \label{heffec}
$$  where  $|B_1 \rangle$,  $|B_2\rangle  \langle  B_2| +  |B_3\rangle
\langle B_3| = |00 \rangle \langle  00| + |11 \rangle \langle 11|$ and
$|B_4 \rangle$  are the  Bell states, as  defined above,  with 
eigenvalues
$\lambda_1  =  0.5$,  $\lambda_2  = \lambda_3=  0.25$  and  $\lambda_4
\approx  0$,  respectively.   Since  the Bell  states  are  completely
entangled, the effective hamiltonian has  no linear terms in the qubit
polarizations and has the form
\begin{equation}
H_{eff}    \sim   \frac{\ln (2)}{2\beta} (1-  (-\sigma_x^{(1)}
\sigma_x^{(2)} + \sigma_y^{(1)} \sigma_y^{(2)} - \sigma_z^{(1)} 
\sigma_z^{(2)}))   +    \frac{\ln (4)}{\beta}
(\frac{1}{2} + 2 \sigma_z^{(1)} \sigma_z^{(2)}), \label{heffec1}
\end{equation}
in  writing  which  the  singlet  term  has been  dropped,  as  it  is
energetically  very far  separated from  the other  three  levels. The
above  analysis  places  in  perspective the  surprising  result  that
although the  system is  evolving, through the  effective Hamiltonian,
the  entanglement density function  remains practically  restricted to
the  3-dimensional subspace,  with a  large contribution  from  a Bell
state,  as the signal 
with the  3-dimensional background acting as
noise. The restriction  of the effective  dynamics from
four to three  levels is also seen in the case  of two qubit evolution
via a dissipative  $S-R$ interaction with a thermal  bath initially at
$T=0$  \cite{ft02},  for the  collective  decoherence model.  However,
there the reason for it is simply given by the fact that for the above
conditions,  the coupling  term connecting  one of  the levels  to the
others goes  to zero,  thereby reducing the  dynamics to  that between
three levels.

An interesting analog  of the discussion in this  Section comes in the
work presented in \cite{hn03}. There  it was shown by the authors that
for  the scenario  where there  exists  a system  consisting of  three
subsystems with  the first and  the third interacting with  each other
via  their  interactions  with   the  mediating  second  subsystem,  a
signature of  entanglement between the first and  the third subsystems
is the  degeneracy in the ground state  of the system. Here  we have a
similar situation with the two qubits interacting with a bath which in
turn  mediates  the   inter-qubit  interaction.   From  our  effective
Hamiltonian $H_{eff}$,  we see that  the first excited state  (not the
ground state),  spanned by  the Bell states  $|B_2 \rangle$  and $|B_3
\rangle$, is degenerate and  the system exhibits a strong entanglement
even  at  finite  temperatures.   Another  work by  the  same  authors
\cite{on02}, studied  the persistence  of mixed state  entanglement at
finite $T$. This would be important as quantum effects can be expected
to dominate in regions where entanglement is nonzero.  They considered
the  transverse Ising  model  and studied  the two-site  entanglement,
using concurrence  as the entanglement measure,  and found appreciable
entanglement in the system at finite $T$ above the ground state energy
gap, one of  their motivations being the influence  of nearby critical
points   to  the   finite  $T$   entanglement.   The   persistence  of
entanglement in  a two-qubit  system interacting with  the bath  via a
purely dephasing interaction (QND) would suggest a broad applicability
of these  concepts, thereby highlighting the  interconnection of ideas
of quantum information to quantum statistical mechanics.


\section{Conclusions}

In  this  article,  we  have   analyzed  in  detail  the  dynamics  of
entanglement in  a two-qubit  system interacting with  its environment
via  a  purely  dephasing  QND  $S-R$  interaction.   The  system  and
reservoir  are initially assumed  to be  separable with  the reservoir
being  in  an initial  squeezed  thermal  state.  Since the  resulting
dynamics becomes mixed, in  order to analyze the ensuing entanglement,
we  have made  use of  a recently  introduced measure  of  mixed state
entanglement via  a PDF.   This enables us  to give a  statistical and
geometrical characterization of entanglement.

After developing  the general dynamics of $N$  qubits interacting with
their bath (reservoir) via a  QND $S-R$ interaction, we specialized to
the  two-qubit case for  applications. Due  to the  position dependent
coupling of the qubits with  the bath, the dynamics could be naturally
divided into  a localized and collective decoherence  regime, where in
the collective decoherence regime, the qubits are close enough to feel
the bath collectively. We analyzed the open system dynamics of the two
qubits, both for  the localized as well as  the collective regimes and
saw that in the collective  regime, there emerges the possibility of a
decoherence-free    subspace    for   the    case    of   zero    bath
squeezing. Interestingly, the dynamics was found to obey a non-trivial
spin-flip  symmetry   operation.  The  existence   of  the  nontrivial
spin-flip symmetry  would explain the emergence  of a decoherence-free
subspace (DFS) \cite{dfs}, thereby  providing a concrete instance of a
DFS. We  made an application of  the two-qubit system  to a simplified
model  of  a  quantum  repeater,  which can  be  adapted  for  quantum
communication over long distances.

We then made  an analysis of the two-qubit  entanglement for different
bath parameters.  We  analyzed both concurrence as well  as the PDF by
finding the  entanglement content of  the various subspaces  that span
the two-qubit Hilbert space.  The analysis of concurrence revealed the
interesting feature of  so called entanglement birth and  death in the
localized  decoherence  model,  while   the  collective  model  saw  a
subsequent revival  of entanglement.  Reservoir squeezing  was seen to
hinder  the   generation  of  entanglement,  though   the  process  of
disentanglement,  as  seen  from   concurrence,  was  robust,  in  the
collective regime, against the effects of squeezed bath.  Although the
PDF agrees  qualitatively in its  predictions with concurrence,  it is
able to extract more information out  of the system as a result of its
statistical-geometrical  nature.  Thus  we  were able  to consider  an
example analogous to NMR  quantum computation, wherein the concurrence
would be zero, and the excess of entangled states over the unpolarized
background is exploited as a resource allowing for non-trivial quantum
information processing.  For the  collective decoherence model the PDF
for the full density  function exhibits a rich entanglement structure,
coming  principally from  the  contributions from  the  one and  three
dimensional projections which carry equal weights thereby suggesting a
tendency of the system to  resist randomization.  This feature is seen
to persist even  for higher temperatures and evolution  times with the
weights of the  subspaces spanned by the four  projection operators of
the  PDF remaining  intact.   The probability  density description  of
entanglement sheds  light on the underlying  dynamics thereby enabling
us  to give  an effective  $T$  dependent dynamics  in the  collective
decoherence  regime.  A  comparison of  this with  some  related works
suggests the  applicability of quantum information  theoretic ideas to
quantum statistical mechanical systems.

\acknowledgments

We wish to thank Shanthanu Bhardwaj for numerical help.


\end{document}